\documentclass[aps,
prl,
reprint,
preprintnumbers,
superscriptaddress,
amsfonts,
amssymb,
amsmath
]{revtex4-2}
\usepackage[T2A]{fontenc}
\usepackage[utf8]{inputenc}
\usepackage{bm,latexsym,mathrsfs,enumerate,bigints}
\usepackage{upgreek}
\usepackage[table,x11names]{xcolor}
\usepackage[breaklinks=true,
unicode=true,
urlcolor = RoyalBlue4,
colorlinks = true,
citecolor = RoyalBlue4,
linkcolor = blue
]{hyperref}

\usepackage{graphicx}
\usepackage{mathtools}
\usepackage{physics,chemformula}
\usepackage[cal=boondoxo,frak=boondox,scr=rsfs]{mathalfa}
\usepackage{savesym}
\usepackage{ulem}
\usepackage[most]{tcolorbox}
\savesymbol{comment}

\usepackage{easyReview}

\usepackage{gitinfo2}
\newcommand{\tn}{T_\textsc{n}}
\renewcommand{\vec}[1]{\mathbf{#1}}

\newcommand{\neel}{N\'{e}el}

\begin{document}
	
	
	\title{Surface-symmetry-driven Dzyaloshinskii--Moriya interaction and canted ferrimagnetism in collinear magnetoelectric antiferromagnet \ch{Cr2O3}}

	\author{Oleksandr~V.~Pylypovskyi}
	\email{o.pylypovskyi@hzdr.de}
	\thanks{These authors contributed equally}
	\affiliation{Helmholtz-Zentrum Dresden-Rossendorf e.V., Institute of Ion Beam Physics and Materials Research, 01328 Dresden, Germany}
	\affiliation{Kyiv Academic University, Kyiv 03142, Ukraine}
	
	\author{Sophie~F.~Weber}
	\email{sophie.weber@mat.ethz.ch}
	\thanks{These authors contributed equally}
	\affiliation{Materials Theory, ETH Z\"{u}rich, Wolfgang-Pauli-Strasse 27, 8093 Z\"{u}rich, Switzerland}
	
	\author{Pavlo~Makushko}
	\affiliation{Helmholtz-Zentrum Dresden-Rossendorf e.V., Institute of Ion Beam Physics and Materials Research, 01328 Dresden, Germany}
	
	\author{Igor~Veremchuk}
	\affiliation{Helmholtz-Zentrum Dresden-Rossendorf e.V., Institute of Ion Beam Physics and Materials Research, 01328 Dresden, Germany}
	
	\author{Nicola~A.~Spaldin}
	\email{nicola.spaldin@mat.ethz.ch}
	\affiliation{Materials Theory, ETH Z\"{u}rich, Wolfgang-Pauli-Strasse 27, 8093 Z\"{u}rich, Switzerland}
	
	\author{Denys~Makarov}
	\email{d.makarov@hzdr.de}
	\affiliation{Helmholtz-Zentrum Dresden-Rossendorf e.V., Institute of Ion Beam Physics and Materials Research, 01328 Dresden, Germany}
	
	\date{October 20, 2023}
	
	\begin{abstract}
		Antiferromagnets are normally thought of as materials with compensated magnetic sublattices. This adds to their technological advantages but complicates readout of the antiferromagnetic state. We demonstrate theoretically the existence of a Dzyaloshinskii--Moriya interaction (DMI) which is determined by the magnetic symmetry classes of \ch{Cr2O3} surfaces with an in-plane magnetic easy axis. The DMI explains a previously predicted out-of-plane magnetization at the nominally compensated surfaces of chromia, leading to a surface-localized canted ferrimagnetism. This is in agreement with magnetotransport measurements and with  density functional theory predictions which further allow us to quantify the strength of DMI. The temperature dependence of the transversal resistance for these planes shows distinct behavior in comparison with that of the \ch{Cr2O3} $c$~plane, which we attribute to the influence of DMI. Our work provides a framework to analyze surface-driven phenomena in antiferromagnets, and motivates the use of nominally compensated chromia surfaces for antiferomagnetic spintronics and magnonics.
	\end{abstract}
	
	\maketitle
	
\begin{figure*}
		\includegraphics[width=\linewidth]{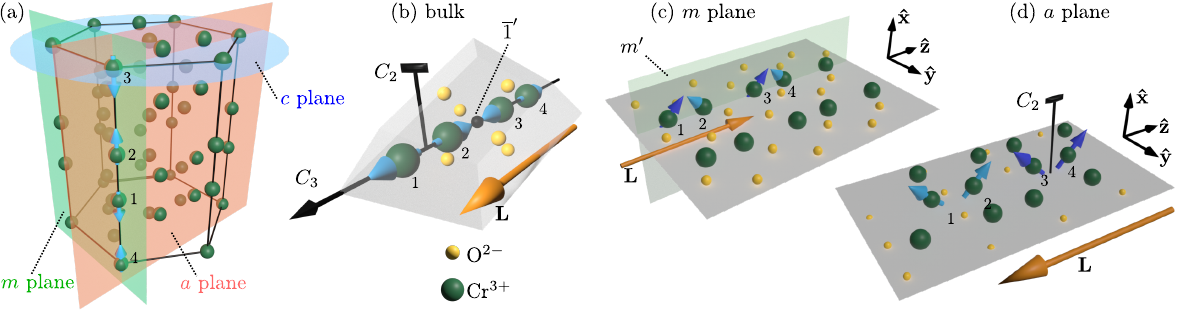}
		\caption{\textbf{Bulk and surface symmetries.} (a)~Cr atoms (green spheres) in the hexagonal cell of \ch{Cr2O3}. $c$~plane $(0\,0\,1)$, $m$~plane $(1\,0\,0)$ and $a$~plane $(\overline{1}\,2\,0)$ are shown by the semitransparent blue, green and red planes, respectively. Blue arrows show directions of magnetic moments for the selected ions. (b)~The basis of atoms in the bulk rhombohedral unit cell of \ch{Cr2O3} with the magnetic ordering $\{+-+-\}$. Here, yellow and green spheres correspond to the~O~and numbered Cr ions, respectively. The yellow arrow indicates direction of the \neel{} vector~$\vec{L}$.  Three-fold and two-fold rotation axes $C_3$ and $C_2$, respectively, are shown, $\overline{1}'$ corresponds to the center of anti-inversion. (c)~Schematic of the $m$~plane with the mirror plane $m'$. Inequivalent moments are shown by bright and dark blue arrows. Axis $\vec{\hat{x}}$ is parallel to the surface normal. (d)~Schematics of the $a$~plane with the two-fold rotation axis $C_2$. Other notations same as in panel (c).
		}
		\label{fig:intro}
	\end{figure*}

	\textit{Introduction.} 
	Chromia \ch{Cr2O3} is a rare example of a room temperature uniaxial magnetoelectric antiferromagnet (AFM) with the bulk \neel{} temperature $\tn = 308$\,K. Beyond its potential in energy efficient AFM-based magnetoelectric data storage~\cite{Kosub17,Mahmood21,Hedrich21}, this material offers a convenient platform for fundamental explorations of spin Hall physics~\cite{Li20d,Makushko22}, THz magnetization dynamics~\cite{Li20d}, spin superfluidity \cite{Yuan18b} and electric-field-only manipulation of magnetism~\cite{Mahmood21}. These studies build on the solid knowledge collected for the highest symmetry surface cut of chromia, terminated by the  $(0\,0\,1)$ surface ($c$~plane)~\cite{Belashchenko10,Weber23}~[Fig.~\ref{fig:intro}(a)]. This surface hosts one AFM sublattice and features substantial magnetization perpendicular to the surface, whose sign is linked to the bulk \neel{} vector $\vec{L}$, proportional to the staggered sum of the individual magnetic moments $\bm{\mu}_{1\ldots4}$ in the unit cell with odd and even indices belong to different AFM sublattices~\cite{Andreev96,Belashchenko10}. This coupling enables all-electric access to the AFM order parameter using conventional magnetotransport methods. In contrast to the $(0\,0\,1)$ surface, the coupling between $\vec{M} \propto \sum_i \bm{\mu}_i$ and $\vec{L}$ is symmetry-forbidden in bulk chromia. Bulk \ch{Cr2O3} does have a symmetry-allowed coupling between the primary order parameter $\vec{L}$ and another antiferromagnetic vector, $\vec{L}_3 \propto \bm{\mu}_1 + \bm{\mu}_2 - \bm{\mu}_3 - \bm{\mu}_4$, due to an antisymmetric Dzyaloshinskii--Moriya interaction (DMI)~\cite{Dzialoshinskii57,Moriya60a}. However, in the easy-axis ground state, $\vec{L}_3 \equiv 0$ and thus the bulk DMI and corresponding coupling vanishes~\cite{Mu19}.
	
	The technological potential of thermodynamically stable chromia surfaces other than the $c$ plane has not been explored, in part due to a lack of understanding of the surface magnetization and its link to the \neel{} vector. In particular, high-symmetry surfaces perpendicular to the $c$~plane such as the $a$~planes and $m$~planes [Fig. \ref{fig:intro}(a)] are magnetically compensated if the magnetic ordering at the surface does not deviate from the bulk. Thus, traditionally, the possibility of surface magnetization on these planes was not considered. However, recent experiments provide evidence for a sizeable spin transport and existence of finite magnetization in particular at the $m$~plane $(1\,0\,0)$~and $a$~plane $(\overline{1}\,2\,0)$ of chromia~\cite{Iino19,Rodriguez22,Erickson23,Du23}, which could be highly relevant for magnonics and physical phenomena such as spin superfluidity because of the in-plane orientation of $\vec{L}$~\cite{Yuan18b}. Furthermore, there is an active search for the uncompensated magnetization, which is coupled to the \neel{} vector~\cite{Makushko22,Du23,Lai23,Weber23a} and signature of DMI in non-collinear textures like domain walls in chromia~\cite{Wornle21}.
	
	This naturally brings attention to crystal surfaces, which are the primary source of  interfacial DMI. In contrast to symmetry breaking at surfaces of ferromagnets, the surface-induced phenomena in AFMs are less well studied. 	Due to the lack of a complete theory of surface magnetism of AFMs, interpretation of magnetotransport, magnonics and spin-texture imaging experiments often interpolates from the behavior of bulk AFMs. This simplification can hide a broad family of fundamental effects. Recently, the surfaces of AFMs have started to attract attention, facilitated by the establishment of a portfolio of techniques that provide access to the relevant physics. These include magnetic circular dichroism~\cite{Du23}, nitrogen vacancy magnetometry~\cite{Casola18,Hedrich21,Huxter22}, magnetoelectric force microscopy~\cite{Schoenherr17}, magnetotransport measurements~\cite{Kosub15,Shiratsuchi21,Mahmood21,Ujimoto23,Mahmood23} as well as sophisticated theoretical methods combining first-principle and model calculations~\cite{Shi09, Mostovoy10, Fechner18, Hedrich21, Makushko22, Weber23}.
	
	Here, we demonstrate that the magnetic symmetry of the nominally compensated surfaces of \ch{Cr2O3}, i.e., $m$~and $a$~planes, provides a sizeable DMI that changes the magnetic ordering at these surfaces and is responsible for the spin canting within the micromagnetic description of AFMs. 
	The DMI is described and quantified using \textit{ab initio} and micromagnetic approaches. Its physics  can be understood in terms of the single-ion and inter-ion anisotropies, as well as antisymmetric exchange. In contrast to the interfacial DMI induced by the inversion symmetry breaking at interfaces~\cite{Bogdanov89,Fert17}, the surface-symmetry-driven DMI relies on the change of the magnetic symmetry point group approaching the sample's surface from the bulk~\footnote{Thus, it takes into account the inversion symmetry break at the sample's interface.}. This DMI couples the primary \neel{} vector $\vec{L}$ with magnetization $\vec{M}$, which is globally symmetry forbidden in bulk \ch{Cr2O3}. The coupling causes $\vec{M}$ to switch with $\vec{L}$, providing all-electric access to $\vec{L}$ for surfaces in which it lies in-plane. The temperature dependence of the transversal resistance shows that the thermodynamic properties of these surfaces differ from those of the $c$~plane, which we attribute to the presence of DMI. Furthermore, we demonstrate that the surface-symmetry-driven DMI results in a change of the magnetic ordering from a collinear bipartite AFM in the bulk to canted ferrimagnetic or 4-sublattice AFM at the surface.
	
	\textit{Single crystal \ch{Cr2O3}.} Bulk chromia belongs to the magnetic symmetry point group $\overline{3}'m'$ that includes 3-fold and 2-fold rotation axes and the center of anti-inversion (i.e., inversion center with time inversion)~[Fig.~\ref{fig:intro}(b)].
	In \ch{Cr2O3}, the bulk energy of the uniform state $\vec{L} = \{L_x, L_y, L_z \}= \text{const}$ is $w_\text{bulk} = \lambda M^2 - K L_z^2$~\cite{Dzialoshinskii57} (supplemental material \footnote{Supplemental material [link provided by publisher] also cites~\cite{Andreev96,Siratori92,Dzialoshinskii57,Marchenko81,Belashchenko10,Weber23,Weber23a,Kresse96,Blochl94,Anisimov97,Dudarev98,Ma15a,Samuelsen70,Shi09,Hedrich21,Kota17a,Veremchuk22a,Parthasarathy19,Hoser12,Mu19,Foner63,Astrov61,Borovik-Romanov06,Morrish95,AlMahdawi17,Gomonay04,Skubic08,Callen66,Tachiki58,Zhang22}.}, Sec.~I), where $\lambda$ is the constant of the uniform exchange, and $K > 0$ is the anisotropy constant. Here and in the following, $\vec{\hat{z}}$ is the direction parallel to the $c$~axis. The high-symmetry $c$-plane cut of chromia has a finite magnetization originating from nonrelativistic exchange, which is proportional to the sublattice magnetization~\cite{Belashchenko10,Weber23}.
	
	To determine the magnetic state of a semi-infinite slab with the surface of a given crystallographic cut, one should complement $w_\text{bulk}$ by the surface energy density~$w_\text{s}$. In the following, we will focus on the $m$~plane  and $a$~plane surfaces, which have an in-plane magnetic easy axis. The allowed components of magnetization on a surface are determined by the subset of bulk symmetry operations that keep the direction of the surface normal invariant~\cite{Marchenko81,Belashchenko10, Weber23a}.
	
	\begin{figure*}
		\includegraphics[width=0.9\linewidth]{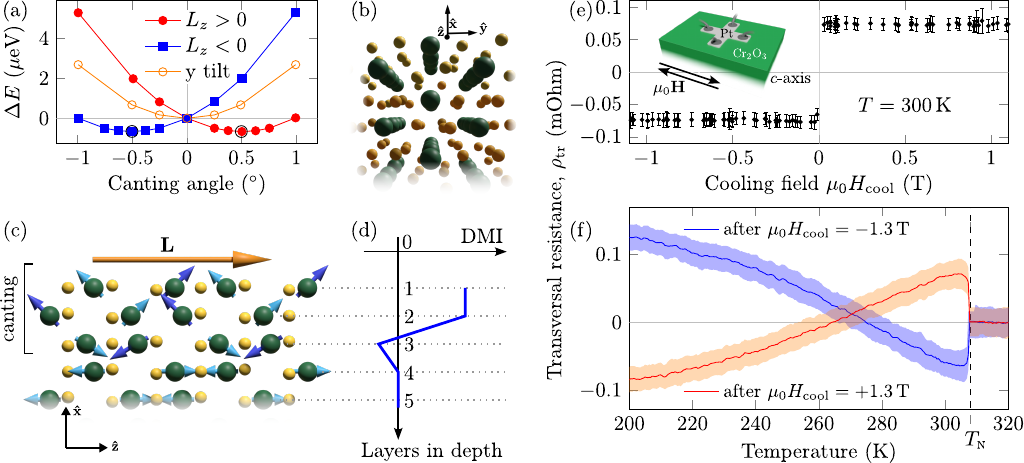}
		\caption{\textbf{$m$-plane \ch{Cr2O3} surface.} (a) Calculated change in energy per surface as a function of canting angle for both directions of the N\'{e}el vector $\vec{L}$. Positive (negative) canting corresponds to an out-of-plane magnetization toward vacuum (bulk). The energy minima for canting along $\vec{\hat{x}}$ are indicated by black circles. (b) Side view of the \ch{Cr2O3} slab in DFT simulations and (c) its front view with arrows indicating the direction of magnetic moments. The N\'{e}el vector $\vec{L}$ is shown by the yellow arrow. (d) Schematics of the change in effective DMI per layer. (e) Experimentally measured spontaneous transversal resistance vs cooling magnetic field. Inset shows schematic of the setup. (f) Measured change of the transversal resistance during zero field warming after cooling in positive and negative fields (red and blue lines, respectively). Fade regions around lines indicate standard deviation of the data. Dashed line indicates the N\'{e}el temperature $T_\textsc{n} = 308$\,K.}
		\label{fig:m-plane}
	\end{figure*}
			
	\textit{Surface magnetism of $m$-plane \ch{Cr2O3}.}  We first discuss the magnetism on the $m$-plane of chromia, which possesses the magnetic symmetry group $m'$ [Fig.~\ref{fig:intro}(c)]. The mirror plane coupled to time-reversal $m'$ transforms $\bm{\mu}_1$ into $\bm{\mu}_3$ ($\bm{\mu}_2$ into $\bm{\mu}_4$) which belong to the same AFM sublattices, This changes the surface symmetry of $m$-plane chromia to that of a \textit{ferrimagnet} rather than of an AFM~\footnote{Mathematical definition of an AFM requires the symmetry operation in magnetic point symmetry group, which transforms sublattices into each other, which makes them distinct from ferrimagnets at the compensation point~\cite{Ivanov05c}}. We determine the surface energy density $w_\text{bulk}$ as the scalar function that is a sum of bilinear and quadratic forms on components of $\vec{M}$ and $\vec{L}$ and is invariant under $m'$. For the $m$~plane it is given by
	\begin{equation}\label{eq:energy-m-plane}
		w_\text{s}^{m} = \lambda_\text{s} M^2 + D(\vec{M}\cdot \vec{L}) + D_{xz}M_xL_z + D_{zx}M_zL_x.
	\end{equation}
	Here, $\vec{\hat{x}}$ is the direction parallel to the surface normal, and $\lambda_\text{s}$ is the constant of the uniform exchange at the surface, which is of the order of $\lambda$. Coefficients $D$, $D_{xz}$ and $D_{zx}$ correspond to the surface-symmetry-driven homogenous Dzyaloshinskii terms which are absent in bulk \ch{Cr2O3} because coupling between $\vec{M}$ and $\vec{L}$ is forbidden for the $\overline{3}'m'$ symmetry group. Their microscopic origin can be determined by considering the spin Hamiltonian with single- and inter-ion anisotropies and antisymmetric exchange~(\cite{Note2}, Sec.~I.C). The term with coefficient $D$ originates from the single-ion anisotropy and is responsible for the emergent ferrimagnetism. Coefficients $D_{xz} = D_\text{sym} + D_\text{asym}$ and $D_{zx}  = D_\text{sym} - D_\text{asym}$ quantify the spin canting at the surface. The symmetric term $D_\text{sym}$ stems from the single-ion anisotropy, while the asymmetric term $D_\text{asym}$ consists of contributions from the inter-ion anisotropy and antisymmetric exchange.
	
	The equilibrium state, determined by the energy minimum for $w_\text{bulk}$ and $w_\text{s}^m$, corresponds to
	\begin{equation}\label{eq:m-ground}
		\vec{L} = L_z\vec{\hat{z}},\quad \vec{M}  = - \dfrac{L_z}{2\lambda_\text{s}} \left\{ D_{xz}, 0, D \right\} + \mathcal{O}(\lambda^{-2}).
	\end{equation}
	Thus, the $m$~plane cut behaves as a canted ferrimagnet with the magnetization in the $xz$ plane uniquely determined by the bulk ground state $\vec{L}$. We stress that the presence of relativistic terms in Eq.~\eqref{eq:energy-m-plane} leading to $\vec{M}\neq 0$ in~\eqref{eq:m-ground} makes the $m$-plane surface qualitatively distinct from the $c$~plane of \ch{Cr2O3}, where the surface magnetization is due to exchange~\cite{Marchenko81,Andreev96,Belashchenko10}~(see also \cite{Note2}, Sec.~I.A).
	
	Constrained magnetic calculations within density functional theory (DFT)~(\cite{Note2}, Sec.~II)~\cite{Weber23a} confirm that the vacuum-terminated surface of the $m$-plane \ch{Cr2O3} having $M_x,M_z\neq0$ is energetically favorable compared to an $m$-plane surface with $M_x=M_z=0$, consistent with a surface magnetization of the form of Eq.~\eqref{eq:m-ground}. We fix \ch{Cr} magnetic moments in the center two layers of a four-layer vacuum-terminated slab of $\mathrm{Cr_2O_3}$ with an $m$-plane surface to lie completely along the bulk \neel{} vector ($[0\,0\,1]$ crystallographic direction) with the ground state magnetic ordering. On the top and bottom outermost surface layers, we induce a surface magnetization by constraining the Cr moments to cant with equal angles along the surface normal, while allowing their magnitudes along $L_z$ to vary [Fig.~\ref{fig:m-plane}(a)]. The canting angle is defined with respect to the $L_z$ $[0\,0\,1]$ direction and we define positive angles as canting toward vacuum, and negative as moments canting toward the bulk. 
		
	\begin{figure}
		\includegraphics[width=0.9\linewidth]{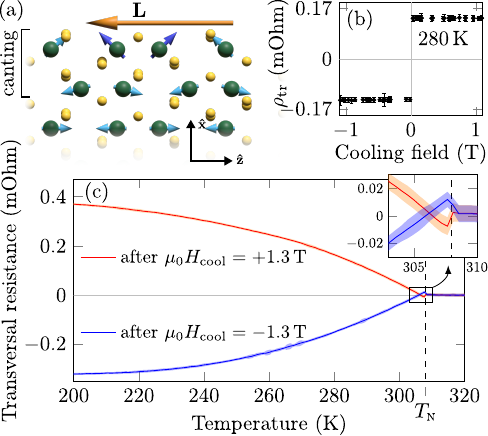}
		\caption{\textbf{$a$-plane \ch{Cr2O3} surface.} (a) Front view of the slab used in DFT simulations with arrows indicating the direction of magnetic moments.   (b) Measured spontaneous transversal resistance vs cooling magnetic field. (c) Change of the transversal resistance measured during zero field warming after cooling in positive and negative fields (red and blue lines, respectively). Fade regions around lines indicate standard deviation of the data. Dashed line indicates the N\'{e}el temperature $T_\textsc{n} = 308$\,K.
		}
		\label{fig:a-plane}
	\end{figure}
	
	In Fig.~\ref{fig:m-plane}(a) we plot the change in total energy per formula unit with respect to energy at $0^{\circ}$ canting ($M_x=0$) as function of the canting angle. We perform two sets of calculations, corresponding to the two bulk AFM domains with $L_z>0$ and $L_z<0$. In line with the symmetry analysis, the DFT data in Fig.~\ref{fig:m-plane}(a) demonstrate that the energy minimum corresponds to a finite canting angle of around $+0.5^{\circ}$ ($-0.5^{\circ}$) for $L_z>0$ ($L_z<0$). The canting results in an induced out-of-plane magnetization of about $+0.1\mu_\textsc{b}$ ($-0.1\mu_\textsc{b}$) for every four $\mathrm{Cr}$ surface moments within the bulk unit cell. The in-plane ferrimagnetic magnetization $M_z$ is smaller than $M_x$ by about a factor of two. Furthermore, we study spin canting in sub-surface layers [Fig.~\ref{fig:m-plane}(b) and \cite{Note2}~Sec.~II]. The second layer of Cr ions reveals roughly the same magnitude of $M_x$ and $M_z$ as the topmost layer while the third layer of ions reveals a small canting of order of $0.1^\circ$ in the opposite direction [Fig.~\ref{fig:m-plane}(c,d)]. 
	
	The values of spin canting obtained from DFT allow us to quantify the DMI and other material parameters in Eq.~\eqref{eq:energy-m-plane} for low temperatures, see Supplementary Table~II~(\cite{Note2}, Sec.~III). 	The degree of ferrimagnetic asymmetry between magnetic sublattices is determined by $D \approx 0.5 \times 10^{-15} \,\text{T$^2$m$^4$/J}$. The inequality of the off-diagonal coefficients $D_{xz} \approx 1\times 10^{-15}\,\text{T$^2$m$^4$/J}$ and $D_{zx} \approx 3\times 10^{-15}\,\text{T$^2$m$^4$/J}$ indicate that both symmetric and antisymmetric components of the DMI are sizeable. The estimated values of DMI can be compared with other materials by normalizing relative to the sublattice magnetization  $M_0 \approx 5\times 10^5 $\,A/m~\cite{Samuelsen70}. Then, $\widetilde{D}_{xz} = D_{xz}/(4M_0^2) \approx 1$\,mJ/m$^2$. This value is of the same order as the interfacial DMI in asymmetric Co sandwiches~\cite{Boulle16,Volkov21} and the values reported for yttrium iron garnets with asymmetric interfaces~\cite{Wang20f}. The physical consequence of the surface-symmetry-driven DMI is the coupling between the surface magnetization and the bulk order parameter.   
    
    To verify {the presence of coupling between $\vec{M}$ and $\vec{L}$}, we perform magnetotransport measurements of a 3-nm-thick Pt thin film prepared on the $m$~plane of a \ch{Cr2O3} single crystal [inset in Fig.~\ref{fig:m-plane}(e), ~\cite{Note2}, Sec.~IV,V]. The sample is cooled from 325\,K (above  $\tn$) to the temperature of interest in an applied magnetic field $\vec{H}_\text{cool}$ oriented along the chromia $c$~axis. This field cooling protocol assures unique selection of a single AFM domain in the sample, fixing the direction of the bulk \neel{} vector $\vec{L}$~\cite{Brown98,Ashida14}.
	Subsequently, the magnetic field is lowered to zero and the transversal resistance is measured at remanence. We determine that the transversal resistance $\rho_\text{tr}$ changes sign with the polarity of the cooling field [Fig.~\ref{fig:m-plane}(e)]. This indicates that the orientation of the out-of-plane component of the surface magnetization $M_x$ at the $m$~plane \ch{Cr2O3} is linked to the bulk $\vec{L}$. Although the physical origin of the out-of-plane magnetization at the surface of $m$~plane is different than that of $c$~plane chromia (relativistic vs. exchange), the experimental fingerprint is similar. Namely, the sign of the transversal resistance $\rho_\text{tr}$ is sensitive to the sign of the interfacial $M_x$~\cite{Kosub15,Iino19}. In the paramagnetic state ($T > \tn$), transversal resistances measured in positive and negative cooling fields are equal, $\rho_\text{tr}^+ = \rho_\text{tr}^-$ [Fig.~\ref{fig:m-plane}(f)]. 
	
	In contrast to the established magnetotransport studies of $c$-plane chromia, where transversal resistance changes monotonically with temperature~\cite{Kosub15,Ji18,Schlitz18,Moriyama20,Wang22d},
	(\cite{Note2}, Sec.~V), the measured $\rho_\text{tr}$ of the $m$-plane sample is not monotonic. In particular, we observe a crossover of the two curves at $T_m \approx 270$\,K which is evidence of the change of the magnetization sign at the surface~[Fig.~\ref{fig:m-plane}(f)]. Our Monte Carlo simulations show that weakening of the exchange bonds at the surface does not lead to the reduction of $\tn$ for the surface layer of moments~(\cite{Note2}, Sec.~VI). Thus, it is unlikely that the reversal of the magnetization sign is related to transition of the first two layers of Cr to a paramagnetic state and domination of the contribution of the third layer of moments which has the opposite direction [Fig.~\ref{fig:m-plane}(c,d)]. We attribute the crossover to the temperature dependence of the DMI, specifically, the different temperature scalings of the single-ion anisotropy determining $D_\text{sym}$ and the antisymmetric exchange with the inter-ion anisotropy determining $D_\text{asym}$~(\cite{Note2}, Sec.~VI).
		
	\textit{Surface magnetism of $a$-plane \ch{Cr2O3}.} We perform the same theoretical and experimental analysis for another nominally magnetically compensated crystallographic cut of \ch{Cr2O3}, the $a$~plane [Fig.~\ref{fig:intro}(d)~(\cite{Note2}, Sec.~I.C)], which belongs to the magnetic point symmetry group $2$ having only the two-fold rotation axis. The surface magnetic energy density determined by the symmetry considerations in the same way as for the $m$~plane is given by
	\begin{equation}\label{eq:energy-a-plane}
		\begin{split}
			w_\text{s}^{a} = \lambda_\text{s} M^2 & + D_{xy} M_xL_y + D_{yx}M_yL_x \\
				& + D_{xz}M_xL_z  + D_{zx}M_zL_x,
			\end{split}
	\end{equation}
	with the corresponding ground state magnetic order given by
	\begin{equation}\label{eq:a-ground}
		\vec{L} = L_z\vec{\hat{z}},\quad \vec{M} = - \left\{ \dfrac{D_{xz}}{2\lambda_\text{s}} L_z, 0, 0 \right\} + \mathcal{O}(\lambda^{-2}),
	\end{equation}
	where the $\vec{\hat{x}}$ axis points along the $a$~plane  surface normal. The 2-fold rotation in the $a$-plane surface point group maps moments $\bm{\mu}_{1}$ and $\bm{\mu}_3$ into opposite AFM sublattices $\bm{\mu}_2$ and $\bm{\mu}_4$. Thus, the $a$~plane surface cut of \ch{Cr2O3} behaves as the four-sublattice weak ferromagnet, while the bulk remains a collinear two-sublattice AFM. The weak ferromagnetism at the surface of the chromia $a$-plane is determined by the surface-symmetry-driven DMI term $D_{xz}$.
	
	In contrast to the $m$~plane where we observed finite spin canting for the top three layers, DFT calculations for the $a$~plane~(\cite{Note2}, Sec.~II) reveal that spin canting is present only for the first two layers of magnetic moments [Fig.~\ref{fig:a-plane}(a)]. 
	The value of spin canting angle allows us to estimate $D_{xz} \approx -0.6\times 10^{-15}\,\text{T$^2$m$^4$/J}$~(\cite{Note2}, Sec.~III). 
	In line with the theoretical predictions, magnetotransport measurements [Fig.~\ref{fig:a-plane}(b,c)]~(\cite{Note2}, Sec.~V) indicate the presence of out-of-plane surface magnetization whose sign is reversed upon reversing $\vec{L}$. The calculated canting angle corresponding to the energy minimum for the $a$~plane is about half that for the $m$~plane. A crucial difference between the $a$~plane and $m$~plane is that the sign of the equilibrium surface magnetization predicted by DFT is opposite for a fixed sign of $\mathbf{L}$ [for the $a$~plane, $L_z<0$ corresponds to positive $M_x$, see Fig. \ref{fig:a-plane}(a)]. This effect is also captured experimentally, reflected in the opposite sign of the transversal resistance measured for the $a$-plane and $m$-plane samples after identical field cooling protocols [c.f. Fig.~\ref{fig:m-plane}(f) and Fig.~\ref{fig:a-plane}(c)]. Similarly to the $m$-plane case, a crossover of the $\rho_\text{tr}^\pm$ curves is observed for the $a$~plane although the crossover temperature, $T_a \approx 306$\,K, is closer to $\tn$ [Fig.~\ref{fig:a-plane}(c)]. The crossover in the $a$-plane temperature dependence suggests that the crossover phenomenon cannot be related to a ferrimagnetic compensation point  because the $a$-plane surface of chromia is not ferrimagnetic.
	
	\textit{Discussion.} The combination of the symmetry analysis and DFT calculations presented here can be straightforwardly extended to other systems and applied for the analysis of non-collinear magnetic textures. The latter involves the additional consideration of inhomogeneous DMI terms including derivatives of the magnetic vectors. The distinction between bulk and surface magnetic states could alter interfacial phenomena such as spin pumping~\cite{Schlauderer19,Liu21,Rodriguez22}. Our findings could be further applied for analysis of the surface responses considering features of multisublattice behavior of AFMs~\cite{Gurevich00} as well as ferrimagnet-specific solitons~\cite{Ivanov19,Galkina22,Zaspel23}.
	
	To summarise, we describe the surface-symmetry-driven DMI at nominally compensated surfaces of chromia. This strong DMI of $\sim 1$\,mJ/m$^2$ causes new physical effects at this otherwise collinear magnetoelectric AFM including (i)~sizeable ($\sim0.5^\circ$) spin canting at the surface, with the direction uniquely determined by the bulk \neel{} vector, (ii)~change of the magnetic ordering at the surface from AFM to ferrimagnetic, and (iii)~peculiar temperature dependence of the magnetotransport.
	
	Our findings provide new insight into the physics of antiferromagnetic material surfaces and expand their technological  potential. In particular, the possibility to assess the \neel{} state of in-plane easy axis AFM surfaces in conventional magnetotransport measurements is crucial for low-energy magnetic data storage~\cite{He10a,Kosub17} and logic~\cite{Manipatruni18} devices. Furthermore, the experimentally demonstrated all-electric readout of the state of these AFMs can link spintronics and AFM magnonics, which relies on the long range propagation of spin waves in AFMs with in-plane easy axes~\cite{Yuan18b,Ross20}. {The unusual temperature dependency of the surface magnetization allows control of the sign of the surface magnetization with heating while keeping the bulk antiferromagnetic state.} Our work positions the $m$~and $a$~planes of chromia as promising materials science platforms for studies of long-range magnon propagation and spin superfluidity. 
		
	\textit{Acknowledgements.} We thank Dr. Tobias Kosub (HZDR Innovation GmbH) for insightful discussions on magnetotransport data of chromia. This work is financed in part via the German Research Foundation (DFG) under Grants No. MA 5144/22-1, MA 5144/24-1, MA 5144/33-1 and and via European Union in the frame of the project REGO (ID:~101070066). S.F.W. and N.A.S. were supported by the ERC under the European Union's Horizon 2020 research and innovation programme with grant No. 810451, and by ETH Z\"{u}rich. Computational resources for the DFT calculations were provided by the Swiss National Supercomputing Centre (CSCS) under project number s1128 and by ETH Z\"{u}rich's EULER cluster.
	
%

\end{document}